\documentclass[pre,preprintnumbers,amsmath,amssymb]{revtex4-1}
\usepackage{graphicx,color}
\usepackage{bm}
\usepackage{epsfig}

\usepackage{lipsum}

\usepackage[FIGTOPCAP,tight,raggedright,nooneline]{subfigure}
\usepackage{float}
\graphicspath{{images/}{plots/}}

\begin{document}

\title{Mean perimeter and area of the convex hull of a planar Brownian motion
in the presence of resetting}

\author{Satya N. Majumdar$^1$, Francesco Mori$^1$, Hendrik Schawe$^{2}$, and Gr\'egory Schehr$^1$}
\address{$^1$ LPTMS, CNRS, Univ.~Paris-Sud, Universit\'e Paris-Saclay, 91405 Orsay, France}
\address{$^2$ Laboratoire de Physique Th\'{e}orique et Mod\'{e}lisation, UMR-8089 CNRS, CY Cergy Paris Universit\'{e}, 95510 Cergy, France}

\begin{abstract}
We compute exactly the mean perimeter and the mean area of the convex hull of a $2$-d Brownian motion
of duration $t$ and diffusion constant $D$, in the presence of resetting to the origin at a
constant rate $r$. We show that
for any $t$, the mean perimeter is given by $\langle L(t)\rangle= 2 \pi \sqrt{\frac{D}{r}}\, f_1(rt)$
and the mean area is given by $\langle A(t) \rangle= 2\pi\frac{D}{r}\, f_2(rt)$ where
the scaling functions $f_1(z)$ and $f_2(z)$ are computed explicitly. For large $t\gg 1/r$,
the mean perimeter grows extremely slowly as $\langle L(t)\rangle \propto \ln (rt)$ with time.
Likewise, the mean area also grows slowly as
$\langle A(t)\rangle \propto \ln^2(rt)$ for $t\gg 1/r$. {Our exact results indicate that the convex hull, in the presence of resetting, approaches a
circular shape at late times.}
Numerical simulations are in perfect agreement with our
analytical predictions.
\end{abstract}
\maketitle

\section{Introduction}

Stochastic processes with resetting have recently emerged as a very active area of research in statistical
physics, due to its numerous applications spanning across interdisciplinary fields ranging from ecology to
computer science -- for a recent review see \cite{Reset_review}. The main idea behind resetting is very simple.
When one is searching for a hidden item, using randomised search algorithms (e.g. diffusive search for food by
animals during their foraging period), it is often advantageous to interrupt the process stochastically at a
constant $r$ and restart the dynamics from the initial position. The effect of resetting has been demonstrated in
a wide variety of problems: diffusive processes such as Brownian motion \cite{EM2011.1,
EM2011.2,WEM2013,MV2013,EM2014,MSS2015,PKE2016,BBR2016,Pal2015,PR2017,Reuveni_16}, random walks and L\'evy flights \cite{KMSS2014,KG2015}, random acceleration process \cite{Singh2020}, active
particles \cite{EM2018,Masoliver2019,KSB2020}, enzymatic reactions \cite{RUK2014,Reuveni_16}, foraging ecology~\cite{BS2014,GGC2019},
active transport in living cells \cite{Bres2020}. Various resetting protocols, some going beyond the canonical Poissonian resetting,
have also been studied \cite{MSS2015_1,Besga2020,NG16,Shkilev17,somrita,Bruyne2020,CS15,BEM2017, FBGM2017, CS2018,
BFGM2019, MBMS2020,GPKP2020,Bres2020_1,Bres2020_2}. Another interesting feature of processes under stochastic resetting is that the
resetting drives the system into a nonequilibrium steady state (NESS). Such NESS have been characterized both for
single particle dynamics \cite{Reset_review,EM2011.1,EMM2013,EM2014,EuMe2016} as well in spatially extended systems such
as fluctuating interfaces \cite{GMS2014,MSS2015,GN2016}, reaction-diffusion systems \cite{DHP2014}, Ising model
with Glauber dynamics \cite{MMS2020}, asymmetric exclusion processes \cite{BKP2019,KN2020,Grange2020}, etc.
Finally, diffusion with different resetting protocols have been realised experimentally in optical tweezers
\cite{Besga2020,TPSRR2020}.

\begin{figure}
    \includegraphics[width=0.6\linewidth]{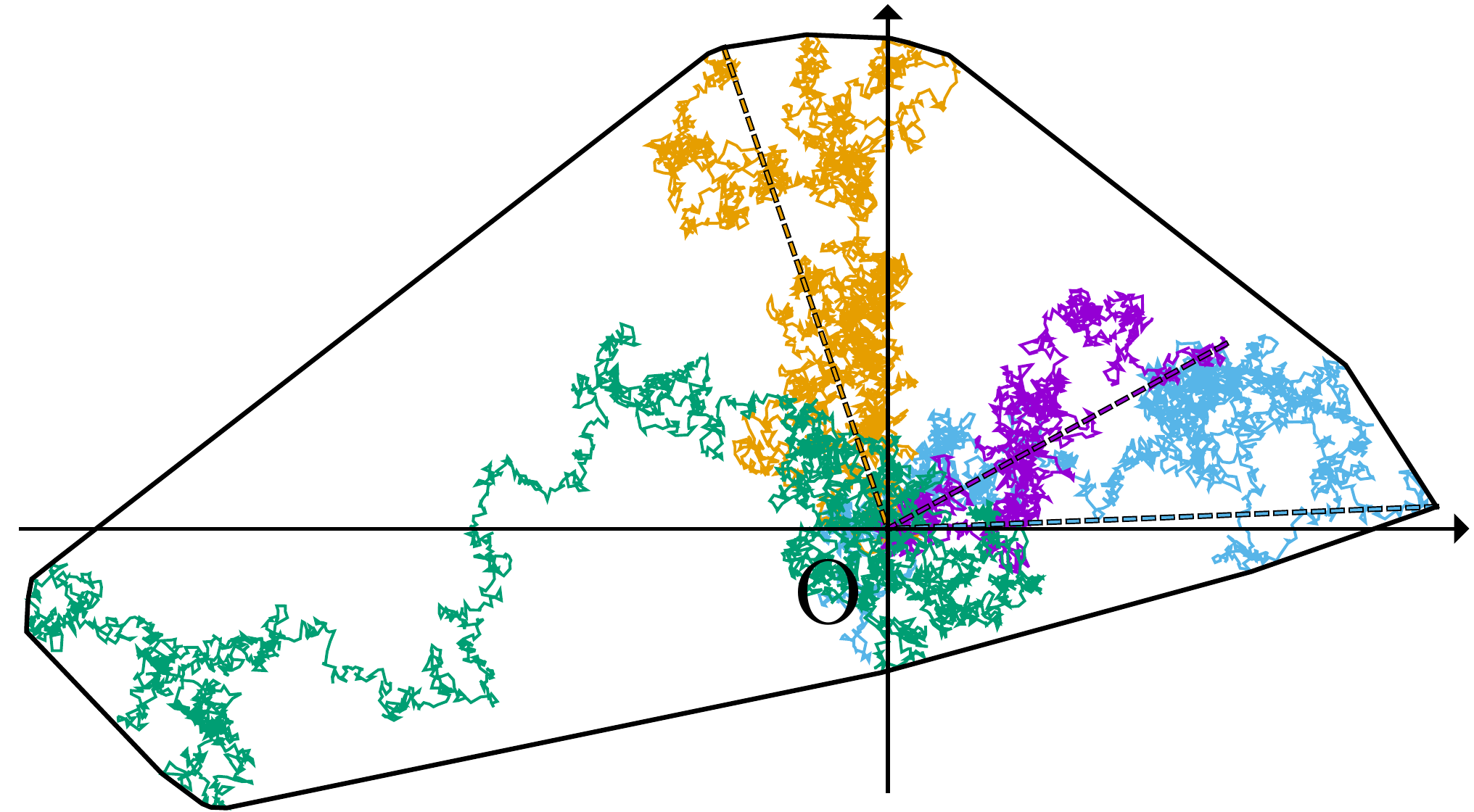}
    \caption{\label{convex_fig}
        Trajectory of a $2$-d RBM (approximated by 15000 discrete steps) where the dashed
        lines indicate the resetting to the origin and the colors distinguish the
        trajectories between the resets for clarity.
        The black polygon represents the convex hull of the trajectory.
    }
\end{figure}

In this paper, we go back to the simplest model, namely the standard Brownian motion with Poissonian resetting to the origin with a constant rate $r$ and we will refer to it as the
reset Brownian motion (RBM). The model is defined more precisely as follows. Let $(x(\tau), y(\tau))$ denote the position of the particle in two-dimensions starting initially
at the origin $(x(0)=0, y(0)=0)$. At any given time $\tau$, the $x$ and the $y$-components are updated as follows
\begin{eqnarray} \label{RBM_rules}
(x(\tau + d\tau),\,  y(\tau + d\tau)) =
\begin{cases}
&(x(\tau) + \eta_x(\tau) d\tau, \, y(\tau) + \eta_y(\tau) d\tau) \;\;\; {\rm with\; proba.} \;\; 1 - r  \,d\tau \;,\\
& \\
&(0,0)  \;\;\; \hspace*{4.2cm}{\rm with\; proba.} \;\; r  \,d\tau \;,
\end{cases}
\end{eqnarray}
where $\eta_x(\tau)$ and $\eta_y(\tau)$ are independent white noises in the $x$ and in the $y$ directions respectively with zero mean and correlations
\begin{eqnarray}
&&\langle \eta_x(\tau) \eta_x(\tau')\rangle = 2 D \delta(\tau-\tau') \label{etaxx_correl} \;, \\
&&\langle \eta_y(\tau) \eta_y(\tau')\rangle = 2 D \delta(\tau-\tau') \label{etayy_correl} \;, \\
&&\langle \eta_x(\tau) \eta_y(\tau')\rangle = 0 \label{etaxy_correl} \;.
\end{eqnarray}
At large times, the RBM in any dimension $d$ reaches a NESS and the stationary position distribution has been computed for all $d$ \cite{EM2014}.
While the position distribution in this NESS has been fully characterized, the spatial structure of the RBM trajectory of a fixed duration $t$ has yet to be characterized. This is particularly relevant in $d=2$ where RBM can be used as a simple model for an animal searching for food starting from its nest at $t=0$ and occasionally going back to the nest to rest. These occasional returns to the nest can be modelled by stochastic resetting moves to the origin, with the assumption that the resetting occurs instantaneously (i.e., the timescale to return is much smaller than the diffusion time scale). How much territory does the animal cover in time $t$? In ecology, this is usually called the home range of the animal \cite{Worton1995}. The trajectory of an animal is usually traced these days using advanced GPS techniques
for animal tracking. A very useful and simple measure of the home range is provided by the convex hull of its trajectory (see Fig.~\ref{convex_fig}) \cite{GAKPY2006}. The statistics of this convex hull, such as
the mean perimeter and the mean area, provide a measure of the geographical territory covered by the animal in time $t$. For an ordinary Brownian motion in the absence of resetting, i.e. when $r=0$, the mean perimeter and the mean area are well known \cite{Takacs1980,ElBachir1983,Letac1993,ch1,ch2}
\begin{eqnarray}
&&\langle L(t) \rangle  = \sqrt{16 \pi D\,t} \;, \label{Lt_BM} \\
&&\langle A(t) \rangle = \pi D\,t \;, \label{At_BM}
\end{eqnarray}
where $D$ is the diffusion constant for each of the $x$ and $y$ coordinates, i.e. $\langle x^2(t)\rangle  = \langle y^2(t)\rangle = 2 D\,t$.

In this paper, our main goal is to investigate how these results (\ref{Lt_BM}) and (\ref{At_BM}) get modified when the resetting rate $r>0$ is switched on, in other words
we want to compute the mean perimeter and the mean area of the $2$-d RBM. Since the position distribution of the $2$-d RBM reaches a stationary state at long times,
one would have perhaps naively guessed that both $\langle L(t) \rangle$ and $\langle A(t) \rangle$ would become time-independent at late times. Our exact computations
show that this naive guess is not correct. The mean perimeter increases very slowly as $\langle L(t) \rangle  \propto \ln(rt)$ at long times $t \gg 1/r$, while the mean area
increases as $\langle A(t) \rangle \propto  \ln^2(rt)$, for $t \gg 1/r$. Actually, in this paper we compute both $\langle L(t) \rangle$ and  $\langle A(t) \rangle$ exactly for all $t$.
Our results can be summarised as follows. We get, for all $t$,
\begin{eqnarray}
&&\langle L(t)\rangle = 2 \pi \sqrt{\frac{D}{r}}\, f_1(rt) \label{mainLt} \\
&&\langle A(t)\rangle = 2 \pi\,{\frac{D}{r}}\, f_2(rt) \label{mainAt}
\end{eqnarray}
where the scaling functions $f_1(z)$ and $f_2(z)$ are given explicitly in Eqs.~(\ref{ex_max.4}) and (\ref{Qz.1}) respectively. Numerical simulations shown in Fig.~\ref{fig_numeric} shows an excellent agreement with our analytical predictions.

The rest of the paper is organised as follows. In Section \ref{Sec:Cauchy} we outline the main method, adapting Cauchy's
formula for closed convex curves in two-dimensions, to compute the mean perimeter and the mean area of the 
$2$-d RBM. In Section \ref{Section:max}, we compute the first two moments of the expected maximum up to time $t$ of the $x$-component of the $2$-d RBM.  In Section \ref{Section:MSD}, we compute the mean square displacement of the $y$-component at the time at which the $x$-component reaches is maximum over the interval $[0,t]$. We then use the exact results from Sections \ref{Section:max}
 and \ref{Section:MSD} to finally compute the mean perimeter and the mean area of the convex hull of 
$2$-d RBM in Section \ref{Sec:final}. Finally, we conclude in Section \ref{Sec:conclusion}. Some details about a Laplace inversion are discussed in Appendix A.

\section{Mean perimeter and mean area of the convex hull of a $2$-d stochastic process via Cauchy's formula}\label{Sec:Cauchy}

Let us briefly recall the key idea developed in Refs.~\cite{ch1,ch2} to compute the mean perimeter and the mean
area of the convex hull of an arbitrary $2$-d stochastic process by adapting Cauchy's formula to a random curve.
This procedure is very general and it essentially maps the problem of computing the mean perimeter and
the mean area of the convex hull of an arbitrary $2$-d process
to the problem of computing the extreme statistics of the one-dimensional component processes in the
$x$ and $y$ directions.

Suppose we have an arbitrary convex domain ${\cal D}$ in two dimensions with
its boundary ${\cal C}$ parametrized as $\{{\cal X}(s),{\cal Y}(s)\}$, where $s$ denotes the arc
distance along the boundary contour ${\cal C}$. According to Cauchy's formula \cite{Cauchy},
the perimeter of the convex domain ${\cal D}$ is given by
\begin{equation}
    L= \int_0^{2\pi} m(\theta)\, d\theta\, ,
    \label{cauchy.1}
\end{equation}
where $m(\theta)$ is known as the support function defined as
\begin{equation}
    m(\theta)= \max\limits_{s}\left[{\cal X}(s)\cos(\theta)+{\cal Y}(s)\sin(\theta)\right]\, .
    \label{support.1}
\end{equation}
The quantity $m(\theta)$ has the simple interpretation: it is the maximum of the projections of all points of the
boundary curve ${\cal C}$ along the
direction $\theta$.

Let us now consider an arbitrary set of $n$
vertices $\{(X_i, Y_i),\, i=1,2,\ldots, n\}$
in $2$-d (e.g., they may represent the positions of a stochastic process in $2$-d at successive times in a given
realization) and construct the convex hull ${\cal C}$ of these vertices. The perimeter of the
convex hull is given by Cauchy's formula in Eq.~(\ref{cauchy.1}). To apply this formula, we need to
first evaluate $\{{\cal X}(s),{\cal Y}(s)\}$ of the convex hull ${\cal C}$ and then compute its maximum over $s$.
This is clearly a difficult problem.
The key observation of Refs.~\cite{ch1,ch2} that bypasses this step is that
the support function $m(\theta)$ of the convex hull can be obtained directly from the
underlying vertices (without the need to first compute $\{{\cal X}(s),{\cal Y}(s)\}$ of ${\cal C}$ and then
maximizing over $s$) as
\begin{equation}
    m(\theta)=  \max\limits_{1\le i\le n}\left[X_i\, \cos(\theta)+ Y_i\, \sin(\theta)\right]\,.
    \label{support.2}
\end{equation}
Next Eq.~(\ref{cauchy.1}) is averaged over all realizations of the stochastic process, i.e., over different realizations of
the vertices $\{(X_i,Y_i)\}$ to get
\begin{equation}
    \langle L_n\rangle = \int_0^{2\pi} \langle m(\theta)\rangle \, d\theta\, .
    \label{support.3}
\end{equation}
Moreover, if the $2$-d process is isotropic (e.g.~the RBM process in $2$-d is isotropic),
$\langle m(\theta)\rangle$ is independent of $\theta$. Consequently, we can just consider the
direction $\theta=0$.
This leads to a simplification of Cauchy's formula as we just need to compute the
expected maximum of the one-dimensional component
process~\cite{ch1,ch2}
\begin{equation}
    \langle L_n\rangle = 2\, \pi\, \langle M_n\rangle\,  \quad {\rm where}\quad M_n= \max\,\left[X_1,X_2,\ldots,X_n\right]\, .
    \label{max.1.1}
\end{equation}

A similar procedure works for the mean area of the convex hull. It has been shown
that the mean area for an arbitrary isotropic $2$-d stochastic process is given by the formula~\cite{ch1,ch2}
\begin{eqnarray} \label{mean_area_Cauchy}
\langle A_n \rangle  = \pi \left[ \langle M_n^2\rangle - \langle y_m^2\rangle(n) \right] \;.
\end{eqnarray}
Here, $\langle M_n^2\rangle$ is the second moment of the maximum $M_n$ of the one-dimensional
$x$ component process as defined in
Eq.~(\ref{max.1.1}). In Eq.~(\ref{mean_area_Cauchy}) the second term refers to the mean square
displacement of the $y$ component
at the time when the $x$ component reaches its maximum.

The results (\ref{max.1.1}) and (\ref{mean_area_Cauchy}) hold for any arbitrary $2$-d isotropic stochastic process
in discrete time,
with $n$ vertices. For a continuous time $2$-d stochastic process, the analogous formulae read
\begin{eqnarray}
&& \langle L(t)\rangle = 2\, \pi\, \langle M(t)\rangle \;, \label{Lt} \\
 && \langle A(t) \rangle  = \pi \left[ \langle M^2(t)\rangle - \langle y_m^2\rangle(t) \right]  \;, \label{At}
\end{eqnarray}
where $t$ denotes the total duration of the process and
\begin{equation}
M(t)= \max[ \{x(\tau)\},\, \forall \,  0\le \tau\le t]\, .
\label{maxt.1}
\end{equation}
Here $x(\tau)$ denotes the continuous time $x$-component process. Similarly, $\langle y_m^2 \rangle (t)$ denotes the mean
square displacement of the continuous time $y$-component process at the time at which the $x$-component $x(\tau)$ reaches
its maximum.

This general procedure has been
successfully used in recent years to compute the mean perimeter and the mean area for several $2$-d
stochastic processes. These include
a single/multiple planar Brownian motions~\cite{ch1,ch2}, planar random acceleration process~\cite{RMS2011},
$2$-d branching Brownian motion with absorption in the context of
epidemic outbreak~\cite{DMRZ2013}, anomalous diffusion processes in $2$-d~\cite{LGE2013}, a $2$-d Brownian motion
confined in the half-plane~\cite{CBM12015,CBM22015}, discrete-time $2$-d random walks, L\'evy flights~\cite{GLM17} and the
run and tumble process in two-dimensions \cite{HMSS_2020}. In this paper, we use these formulae in (\ref{Lt}) and (\ref{At}) to compute exactly the
mean perimeter and the mean area of the convex hull of the $2$-d RBM.

\section{The statistics of the maximum of the $1$-d Brownian motion with resetting}\label{Section:max}

Consider the $2$-d RBM with a constant resetting rate $r$.
The process starts at the origin in the $2$-d plane at time $\tau=0$.
Let
$x(\tau)$ denote the $x$ component of this $2$-d reset Brownian motion. Since for a $2$-d RBM,
the $x$ and the $y$ components evolve independently between resets, the
process $x(\tau)$ just represents a one dimensional Brownian motion, starting
at $x(0)=0$, with resetting to the
$x=0$ at rate $r$. In this section, we compute the statistics of the maximum $M(t)$ of
this one dimensional RBM process $x(\tau)$ up to time $t$. As explained earlier,
the first two moments $\langle M(t)\rangle$ and $\langle M^2(t)\rangle$ are needed as inputs
in the computation of the mean perimeter and the mean area of the convex hull of the $2$-d
RBM. In fact, the distribution of the maximum $M(t)$ for a one dimensional RBM
was already studied in Ref.~\cite{EM2011.1} in a slightly different language and context.
Here we revisit this problem and present a full derivation of the first two moments
explicitly which are needed for our purpose to compute the statistics of the convex hull of the $2$-d problem.

\begin{figure}
    \center
    \includegraphics[width=0.75\linewidth]{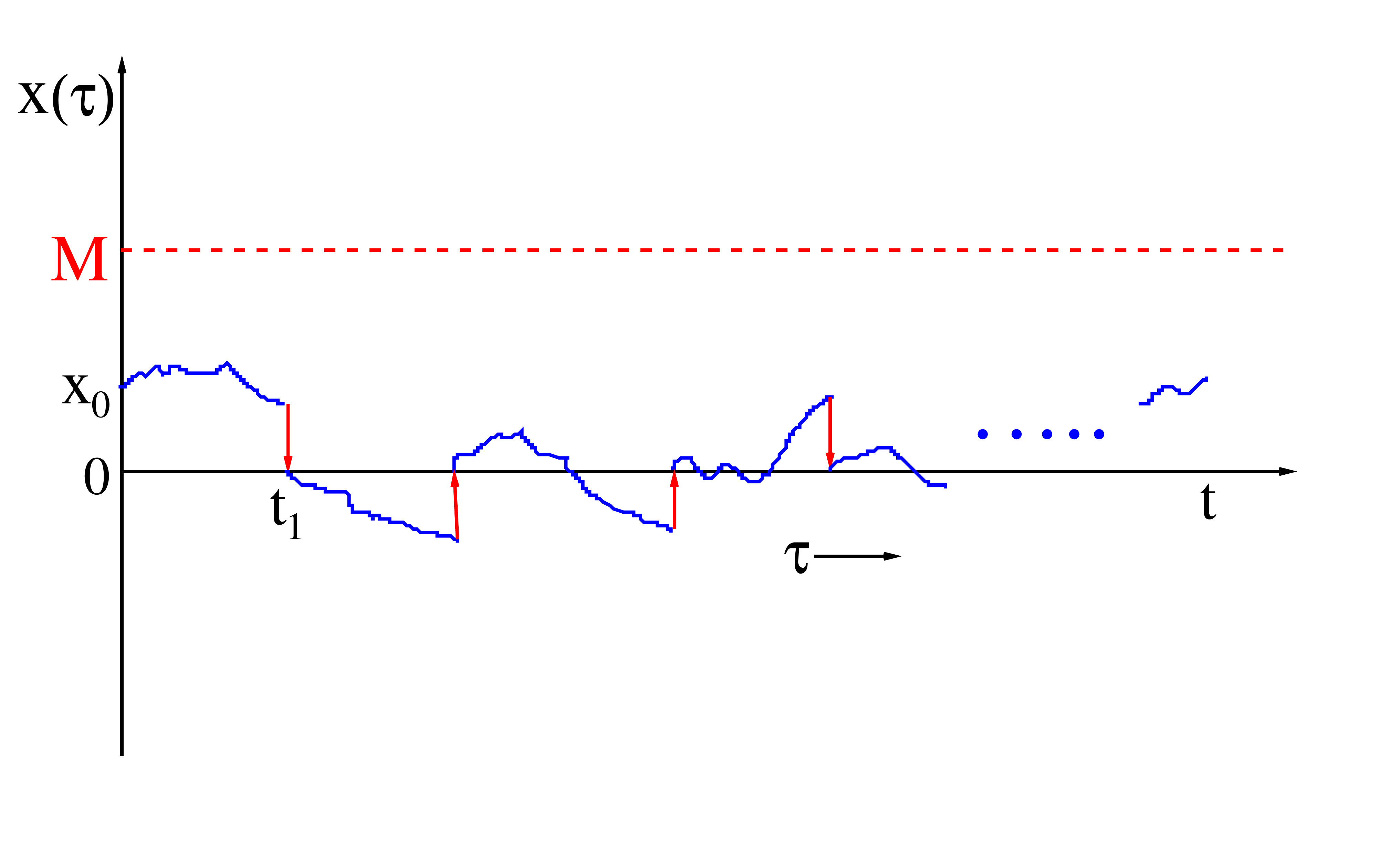}
    \caption{\label{fig_1drbm}
A typical trajectory (schematic) of a $1$-d Brownian motion $x(\tau)$ vs. $\tau$,
with resetting to $x=0$ at a constant rate $r$. The process starts at $x(0)=x_0$,
stays below the level $M$ (with $M\ge x_0$ as shown by the dashed (red) line) up to time $t$.
The resetting events to the origin are shown by directed (red) arrows, with
the first resetting at time $t_1$.}
\end{figure}

Let $M(t)$ denote the maximum of the process $x(\tau)$, starting at $x(0)=0$, up to time $t$
\begin{equation}
M(t)= \max[ \{x(\tau)\},\, \forall \,  0\le \tau\le t]\, .
\label{max.1}
\end{equation}
Clearly $M(t)\ge 0$ since the process starts at the origin. To compute the statistics
of $M(t)$, it is convenient to consider the cumulative distribution of $M(t)$
\begin{equation}
Q_r(M,t)= {\rm Prob.}[M(t)\le M]\, ,
\label{cumul_max.1}
\end{equation}
where the subscript $r$ refers to the process with resetting rate $r$. Let us first define
a more general quantity (that will be useful in the next section also) $S_r(x_0,t|M)$ which
denotes the probability that the process $x(\tau)$, starting at $x_0$, stays below the level
$M$ up to time $t$. Thus, this is just the survival probability of the reset Brownian process
starting at $x_0$, with an absorbing boundary at $x=M$ with $M\ge x_0$ (see Fig.~\ref{fig_1drbm}
for a typical trajectory)~\cite{EVS_review}. If we know the survival probability $S_r(x_0,t|M)$ for all $x_0$,
the cumulative distribution of the maximum can be simply obtained by setting $x_0=0$,
\begin{equation}
Q_r(M,t)= S_r(0,t|M)\, .
\label{cum_surv.1}
\end{equation}

To compute the survival probability $S_r(x_0,t|M)$ we use a
simple renewal argument~\cite{Reset_review} that expresses the survival probability of the reset
process in terms of the survival
probability $S_0(x_0,t|M)$ of the process without reset. Indeed, there are two possibilities:
no resetting or at  least one resetting in the interval $[0,t]$.
If there is no resetting, the survival probability is simply $e^{-r t}\, S_0(x_0,t|M)$.
In the latter case, let $t_1$ denotes the time at which the first resetting occurs. Then
the process renews at $t_1$, starting from the origin. Adding up the two contributions we get
\begin{equation}
S_r(x_0,t|M)= e^{-rt}\, S_0(x_0,t|M)+ r\, \int_0^t dt_1\, e^{-r t_1} \,
S_0(x_0,t_1|M)\, S_r(0,t-t_1|M)
\label{renewal.1}
\end{equation}
The first term corresponds to survival of the process with no resetting in the interval $[0,t]$.
In the second term, $r e^{-r t_1}$ denotes the probability that the first resetting occurs
at $t_1$. The factor $S_0(x_0,t_1|M)$ is the survival probability during this interval $[0,t_1]$,
while $S_r(0,t-t_1|M)$ denotes the survival probability during the time interval $[t_1, t]$.
Note that in the second interval, the subscript $r$ shows that the process is with resetting.
It is convenient to take the Laplace transform of Eq.~(\ref{renewal.1}) so that the
convolution structure in the second term on the right hand side (rhs) can be exploited.
Defining
\begin{equation}
{\tilde S}_r(x_0,s|M)= \int_0^{\infty} S_r(x_0,t|M)\, e^{-s t}\, dt \, ,
\label{laplace_def.1}
\end{equation}
we get from Eq.~(\ref{renewal.1})
\begin{equation}
{\tilde S}_r(x_0,s|M)= {\tilde S}_0(x_0,r+s|M)+ r\, {\tilde S}_0(x_0, r+s|M)\, {\tilde S}_r(0,s|M)\,.
\label{renewal.2}
\end{equation}
Setting $x_0=0$, we obtain
\begin{equation}
{\tilde S}_r(0,s|M)= \frac{ {\tilde S}_0(0,r+s|M)}{1- r\, {\tilde S}_0(0,r+s|M)}\, .
\label{sr0.1}
\end{equation}
Hence, from Eq.~(\ref{renewal.2})
\begin{equation}
{\tilde S}_r(x_0,s|M)=  \frac{{\tilde S}_0(x_0,r+s|M)}{1- r\, {\tilde S}_0(0,r+s|M)}\, .
\label{srx0.1}
\end{equation}

The survival probability $S_0(x_0, t|M)$ of an ordinary Brownian motion (with diffusion constant $D$)
starting at $x_0$
and with an absorbing boundary at $M$ can be very easily computed using the method of
images~\cite{Redner_Book}. Equivalently, its Laplace transform can be computed directly
by solving a backward Fokker-Planck equation~\cite{BF_2005,Persistence_review}.
Indeed, $S_0(x_0,t|M)$ satisfies the backward Fokker-Planck equation
\begin{equation}
\frac{\partial S_0}{\partial t}= D\, \frac{\partial^2 S_0}{\partial x_0^2}
\label{bfp.1}
\end{equation}
in the region $x_0\in [-\infty,M]$ with absorbing boundary condition at $x_0=M$, i.e,
$S_0(x_0=M,t|M)=0$ and $S_0(x_0\to -\infty,t|M)=1$ for all $t$. The last condition
follows from the fact
that if the particle starts at $-\infty$, it definitely stays below $M$ for any finite $t$.
The initial condition is $S_0(x_0,0|M)=1$ for all $x_0<M$.
Taking the Laplace transform of Eq.~(\ref{bfp.1}) with respect to $t$ and using the initial
condition gives
\begin{equation}
-1+ s\, {\tilde S}_0(x_0,s|M)= D\, \frac{d^2 {\tilde S}_0}{dx_0^2} \, .
\label{qtilde.1}
\end{equation}
The solution, using
the two boundary conditions, reads for $x_0\le M$
\begin{equation}
{\tilde S}_0(x_0,s|M)=\frac{1}{s}\,\left[1- e^{-\sqrt{\frac{s}{D}}\, (M-x_0)}\right]\, .
\label{surv_x0.1}
\end{equation}
This Laplace transform can be explicitly inverted to give
\begin{equation}
S_0(x_0,t|M)= {\rm erf}\left(\frac{M-x_0}{\sqrt{4 Dt}}\right)\, ; \quad\,\,
{\rm erf}(z)= \frac{2}{\sqrt{\pi}}\, \int_{0}^{z} e^{-u^2}\, du \, ,
\label{surv_prob_bm.1}
\end{equation}
which coincides, as expected, with the survival probability up to $t$ of an ordinary Brownian motion, starting
at $M-x_0\ge 0$ and with an absorbing boundary at the origin~\cite{Redner_Book,Persistence_review}.
However, for our purpose, the Laplace transform of this survival probability
in Eq.~(\ref{surv_x0.1}) is more useful.

Indeed, substituting the result from Eq.~(\ref{surv_x0.1}) on the rhs of Eq.~(\ref{srx0.1}) gives our desired quantity
\begin{equation}
{\tilde S}_r(x_0,s|M)= \frac{1- e^{-\sqrt{\frac{(r+s)}{D}}\,(M-x_0)}}{s+r\, e^{-\sqrt{\frac{(r+s)}{D}}\, M}}\, .
\label{srx0.2}
\end{equation}
Setting $x_0=0$ in Eq.~(\ref{srx0.2}) and using Eq.~(\ref{cum_surv.1}) we get the
Laplace transform of the cumulative distribution of the maximum (for the process
starting at the origin)
\begin{equation}
{\tilde Q}_r(M,s)= \int_0^{\infty} Q_r(M,t)\, e^{-s t}\, dt=
\frac{1- e^{-\sqrt{\frac{(r+s)}{D}}\, M}}{s+r e^{-\sqrt{\frac{(r+s)}{D}}\, M}}\, .
\label{cum_maxr.1}
\end{equation}
The probability density function (PDF) $P_r(M,t)$ of the maximum $M(t)$ can be obtained from
the cumulative distribution by just taking a derivative, $P_r(M,t)= \partial_M Q_r(M.t)$.
Hence, its Laplace transform, obtained by taking derivative of Eq.~(\ref{cum_maxr.1}), reads
\begin{equation}
{\tilde P}_r(M,s)= \int_0^{\infty} P_r(M,t)\, e^{-s t}\, dt= \frac{(r+s)^{3/2}}{\sqrt{D}}\,
\frac{e^{-\sqrt{\frac{(r+s)}{D}}\, M}}{\left(s+r\, e^{-\sqrt{\frac{(r+s)}{D}}\, M}\right)^2}\, .
\label{pdf_maxr.1}
\end{equation}
It is difficult to invert this Laplace transform explicitly, except at late times \cite{EM2011.1}
when the PDF $P_r(M,t)$ can be shown to converge to the Gumbel law~\cite{EM2011.1}.
For our purpose, namely to compute the mean perimeter and the mean area of the convex hull of the
$2$-d RBM,
we only need the first two moments. The first moment $\langle M(t)\rangle$ was already computed explicitly for
all $t$~\cite{EM2011.1} which we reproduce below for the sake of completeness.
Here we show that the second moment $\langle M^2(t)\rangle$ can also be computed explicitly for all $t$.

\subsection{Expected maximum $\langle M(t)\rangle $}

The Laplace transform of the expected maximum is given by
\begin{equation}
\int_0^{\infty} \langle M(t)\rangle\, e^{-s t}\, dt= \int_0^\infty M\, {\tilde P}_r(M,s)\, dM\, .
\label{ex_max.1}
\end{equation}
Using the explicit result in Eq.~(\ref{pdf_maxr.1}) and performing the integral gives
\begin{equation}
\int_0^{\infty} \langle M(t)\rangle\, e^{-s t}\, dt= \frac{\sqrt{D\,(r+s)}}{r\, s}\, \ln \left(\frac{r+s}{s}\right)\, .
\label{lt_max.1}
\end{equation}
Fortunately, this Laplace transform can be explicitly inverted~\cite{EM2011.1} and
we present its derivation in Appendix A. This gives
\begin{equation}
\langle M(t)\rangle= \sqrt{\frac{D}{r}}\, f_1(r\,t)\, ; \quad {\rm where}\quad
f_1(z)= \int_0^z \frac{dy}{y}\, (1-e^{-y})\, \left[ \frac{e^{-(z-y)}}{\sqrt{\pi\, (z-y)}}+
{\rm erf}\left(\sqrt{z-y}\right)\right]\, ,
\label{ex_max.4}
\end{equation}
where ${\rm erf}(z)= (2/\sqrt{\pi})\, \int_0^z e^{-u^2}\, du$ is the error function.
The asymptotic behaviors of the scaling function $f_1(z)$,
for small and large $z$, can be easily derived. One gets to leading orders~\cite{EM2011.1}
\begin{eqnarray}
f_1(z) \approx \begin{cases}
\frac{2}{\sqrt{\pi}}\, \sqrt{z}  {- \frac{1}{45 \sqrt{\pi}}z^{5/2} } \quad\, {\rm as}\quad z\to 0 \\
\\
\ln z + \gamma_E  \quad \;\;\; \hspace*{1.1cm}{\rm as}\quad z\to \infty \;,
\end{cases}
\label{f1z_asymp}
\end{eqnarray}
where $\gamma_E=0.57721\ldots$ is the Euler constant.

The asymptotic behaviors of $\langle M(t)\rangle$ for $t\ll 1/r$ and $t\gg 1/r$ then follow readily.
One gets~\cite{EM2011.1}
\begin{eqnarray}
\langle M(t) \rangle \approx \begin{cases}
\sqrt{Dt} \left[ \sqrt{\frac{4}{\pi}} {- \frac{1}{45\sqrt{\pi}}(rt)^2}\right]\,  \quad\quad\quad {\rm for}\quad t\ll \frac{1}{r} \\
\\
\sqrt{\frac{D}{r}}\, \left[\ln (rt) + \gamma_E\right]  \quad\hspace*{1.6cm} {\rm for}\quad t\gg \frac{1}{r} \;.
\end{cases}
\label{exmax_asymp}
\end{eqnarray}
The leading term in the first line can be simply understood as follows.
Since $1/r$ is the typical time for a resetting even to take place, when $t \ll 1/r$ the particle has
hardly undergone any resetting and hence
it behaves as a free Brownian motion. Indeed, the leading term in the first line
of Eq.~(\ref{exmax_asymp}) corresponds exactly to the expected maxima of an ordinary Brownian motion~\cite{EVS_review}.
The result for large $t\gg 1/r$ is more interesting. As was shown in Refs. \cite{EM2011.1,MSS2015}, the
RBM approaches a stationary state when $t\gg 1/r$ due to the repeated resettings. Nevertheless,
the expected maximum up to time $t$ still increases with time, albeit very slowly as a logarithm. This
can be understood heuristically using the theory of extreme value statistics (EVS) of weakly correlated
variables~\cite{EM2011.1,Reset_review,EVS_review}. The scaling function $f_1(z)$ describes the full crossover from the
early time Brownian growth of the expected maximum to the late time logarithmic growth.

\begin{figure}
    \center
    \includegraphics[scale=1]{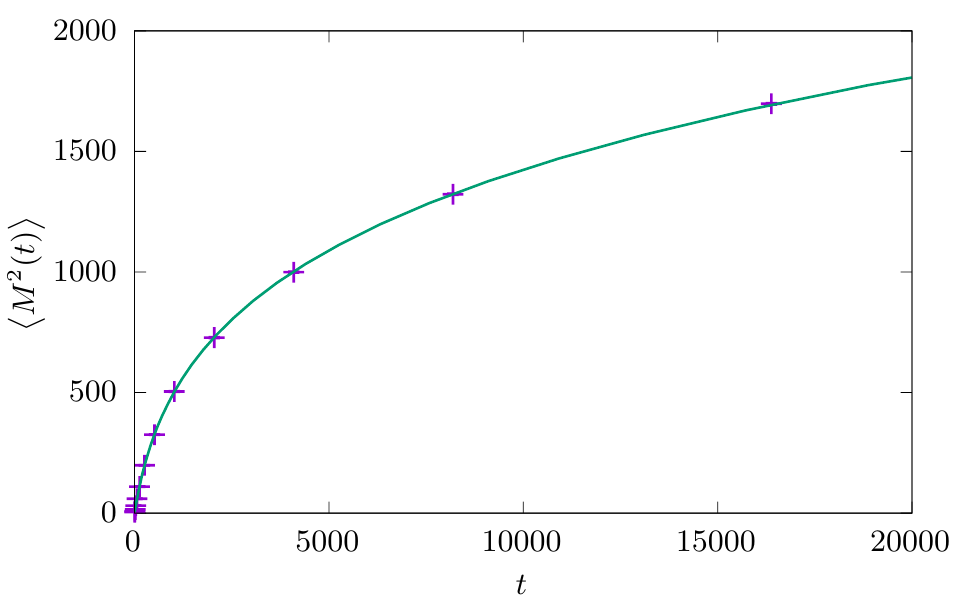}
    \caption{\label{fig_m2}
The second moment $\langle M^2(t)\rangle$ plotted as a function of time $t$.
The solid line represents the exact result $\langle M^2(t)\rangle= (2D/r)\, H(rt)$ with $H(z)$ given in Eq.~(\ref{m2.3}),
while the symbols represent the results from numerical simulations. {Here $D=1/2$ and $r=10^{-2}$.}}
\end{figure}

\subsection{The second moment $\langle M^2(t)\rangle$}
The Laplace transform of the second moment is given by
\begin{equation}
\int_0^{\infty} \langle M^2(t)\rangle\, e^{-st}\, dt= \int_0^{\infty} M^2\, {\tilde P}_r(M,s)\, dM \, .
\label{m2.1}
\end{equation}
Using the result for ${\tilde P}_r(M,s)$ from Eq.~(\ref{pdf_maxr.1}) and carrying out integration we get
\begin{equation}
\int_0^{\infty} \langle M^2(t)\rangle\, e^{-st}\, dt= -\frac{2D}{r\,s}\, {\rm Li}_2\left(-\frac{r}{s}\right)\, ; \quad {\rm where}
\quad  {\rm Li}_2(z)= \sum_{k=1}^{\infty} \frac{z^k}{k^2}\, .
\label{m2.2}
\end{equation}
Substituting the series expansion of the PolyLog function ${\rm Li}_2(-r/s)$ we can then invert the Laplace transform
term by term, by using ${\cal L}_s^{-1}\left(s^{-(k+1)}\right)= t^k/k!$, where ${\cal L}_s^{-1}$ denotes the Laplace
inverse with respect to $s$. This gives
\begin{equation}
\langle M^2(t)\rangle= \frac{2D}{r}\, H(r\,t)
\label{Hz_def}
\end{equation}
where the scaling function $H(z)$ is given by
\begin{equation}
H(z)= \sum_{n=1}^{\infty} \frac{(-1)^{n+1}}{n^2}\, \frac{z^n}{n!}= z\, _3F_3
\left[\{1,1,1\},\{2,2,2\}; -z\right]\, ,
\label{m2.3}
\end{equation}
where $_pF_q$ is the generalised hypergeometric series~\cite{GR}.
The asymptotic behaviors of the scaling function $H(z)$ are given by
\begin{eqnarray}
H(z) \approx \begin{cases}
z- \frac{z^2}{8}  \quad\, {\rm as}\quad z\to 0 \\
\\
\frac{1}{12}\left[6\, \ln^2 z + 12\, \gamma_E\, \ln z + 6\, \gamma_E^2+ \pi^2\right]  \quad {\rm as}\quad z\to \infty
\end{cases}
\label{Hz_asymp}
\end{eqnarray}
This then gives the asymptotic behaviors of $\langle M^2(t)\rangle$ from Eq.~(\ref{m2.2}). Once again, for
$t\ll 1/r$, one recovers the Brownian result, $\langle M^2(t)\rangle \approx 2\,D\,t$ as expected, while
for large $t\gg 1/r$, it grows as $\langle M^2(t)\rangle \approx (D/r)\, \ln^2(rt)$ to leading order.
In Fig.~\ref{fig_m2} we compare our analytical prediction for $\langle M^2(t)\rangle$ in Eqs.~(\ref{m2.2})
along with Eq.~(\ref{m2.3}) with numerical simulations, finding excellent agreement for all $t$.

\section{Computation of the second moment $\langle y_m^2\rangle(t)$}\label{Section:MSD}

In this section, we compute another ingredient needed for the calculation of the
mean area of the convex hull. Let $\{x(\tau),y(\tau)\}$ denote the $x$ and the $y$ component process of the $2$-d Brownian motion
starting at the origin $\{x(0)=0, y(0)=0\}$ and resetting to the origin. When the resetting occurs,
both the $x$ and the $y$ component are reset simultaneously to $x(0)=0$ and $y(0)=0$ respectively.
Thus resetting makes the two component processes highly correlated. Let $t_m$ denote the time
at which the $x$-component achieves its maximum value, say $M$, in the interval $[0,t]$. For the
computation of the mean area, we need to compute $\langle y_m^2\rangle(t)$ where
$y_m= y(\tau=t_m)$ denotes the value of the $y$ component exactly at the
time $\tau=t_m$ when
the $x$ component achieves its maximum in the interval $[0,t]$.
Let us briefly recall how to compute
this for a standard $2$-d Brownian motion in the absence of resetting, i.e., when the reset rate
$r=0$~\cite{ch1,ch2}. In this case, since there is no correlation between the
$x$ and the $y$ component with each of them performing independent Brownian motion, it follows
that $\langle y^2(\tau)\rangle = 2\, D\, \tau$ for any $\tau$. Hence,
in this case $\langle y_m^2\rangle(t)= 2\,D\, \langle t_m\rangle= D\, t$ where one uses $\langle t_m\rangle=t/2$
for the $x$ component which is just a one dimensional free Brownian motion. However, this simple
argument does not work in the presence of a finite resetting rate $r>0$ since the two components get strongly
correlated via the resetting events. Nevertheless, $\langle y_m^2\rangle(t)$
can be computed explicitly
for all $t$ as we show in this section.

Before starting the computation for $\langle y_m^2\rangle(t)$
for the Brownian motion with resetting, let us first
briefly recall some basic facts about the Brownian motion
in the absence of resetting ($r=0$). Consider first
the Green's function or the propagator for a $2$-d Brownian motion.
This is just the probability density $G_0(x,y;\tau)$ that the Brownian motion, starting at the origin $(0,0)$ reaches $(x,y)$
at time $\tau$ and is simply given by the product of two one dimensional propagators
\begin{equation}
G_0(x,y,\tau)= \frac{1}{4\,\pi\, D\, \tau}\, e^{- \frac{x^2+y^2}{4\, D\, \tau}}\, .
\label{G0.1}
\end{equation}
If however one puts an absorbing boundary at $x=M$, then the constrained
propagator of the same process to reach $(x\le M,y)$ at time $\tau$,
while staying below the level $M$ during the whole interval
$[0,\tau]$, can be easily obtained using the method
of images~\cite{Redner_Book,BF_2005}. One gets
\begin{equation}
G_0(x,y,\tau|M)= \frac{1}{4\,\pi\, D\, \tau}\, e^{-\frac{y^2}{4\,D\,\tau}}\,
\left[ e^{-\frac{x^2}{4\,D\, \tau}}- e^{-\frac{(2M-x)^2}{4\,D\,\tau}}\right]\, .
\label{G0_M.1}
\end{equation}
We will need this result shortly.

\begin{figure*}[bhtp]
        \centering
        \subfigure[\label{fig_xt}]{
            \includegraphics[scale=0.2]{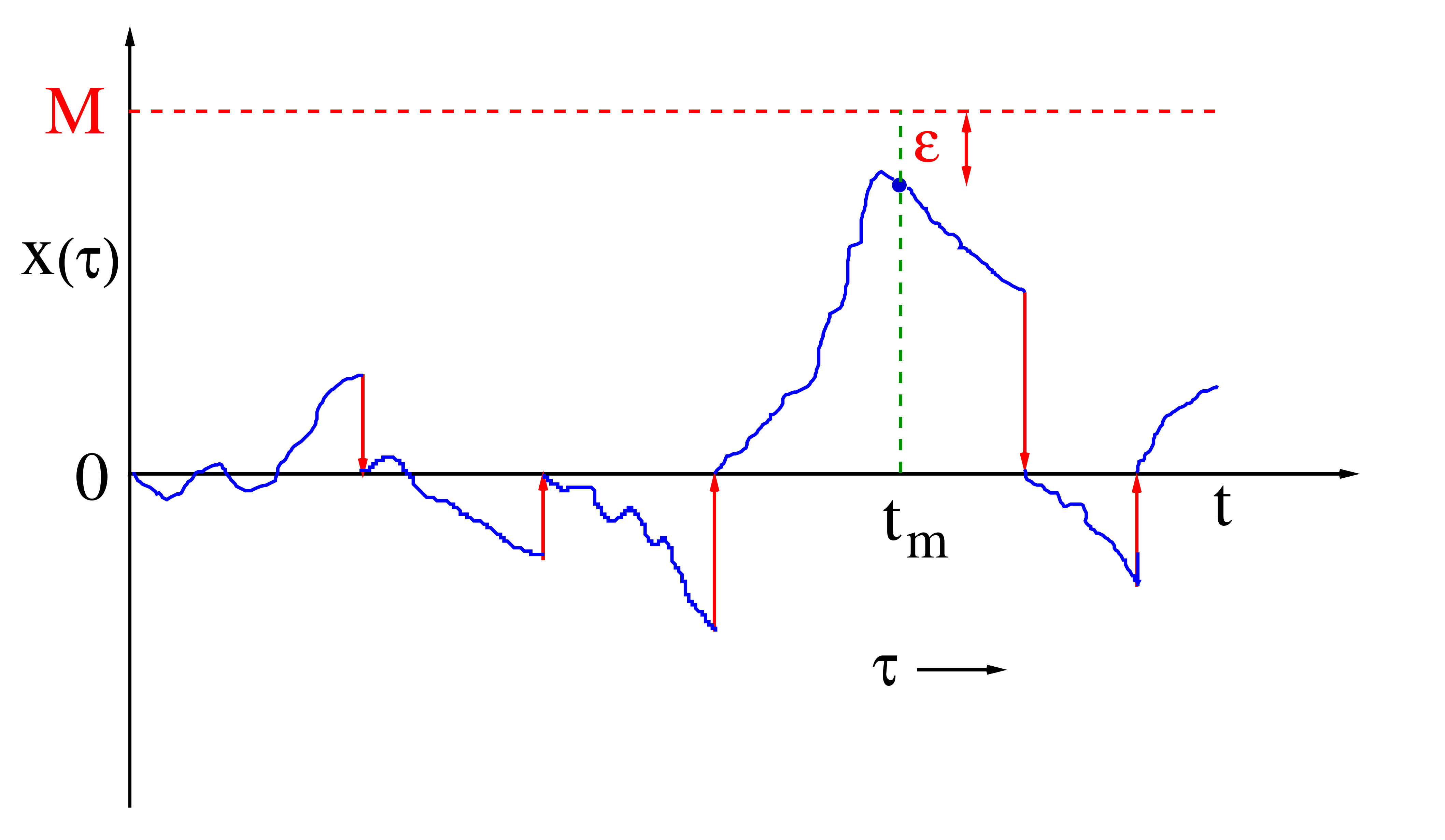}
        }
        \subfigure[\label{fig_yt}]{
            \includegraphics[scale=0.2]{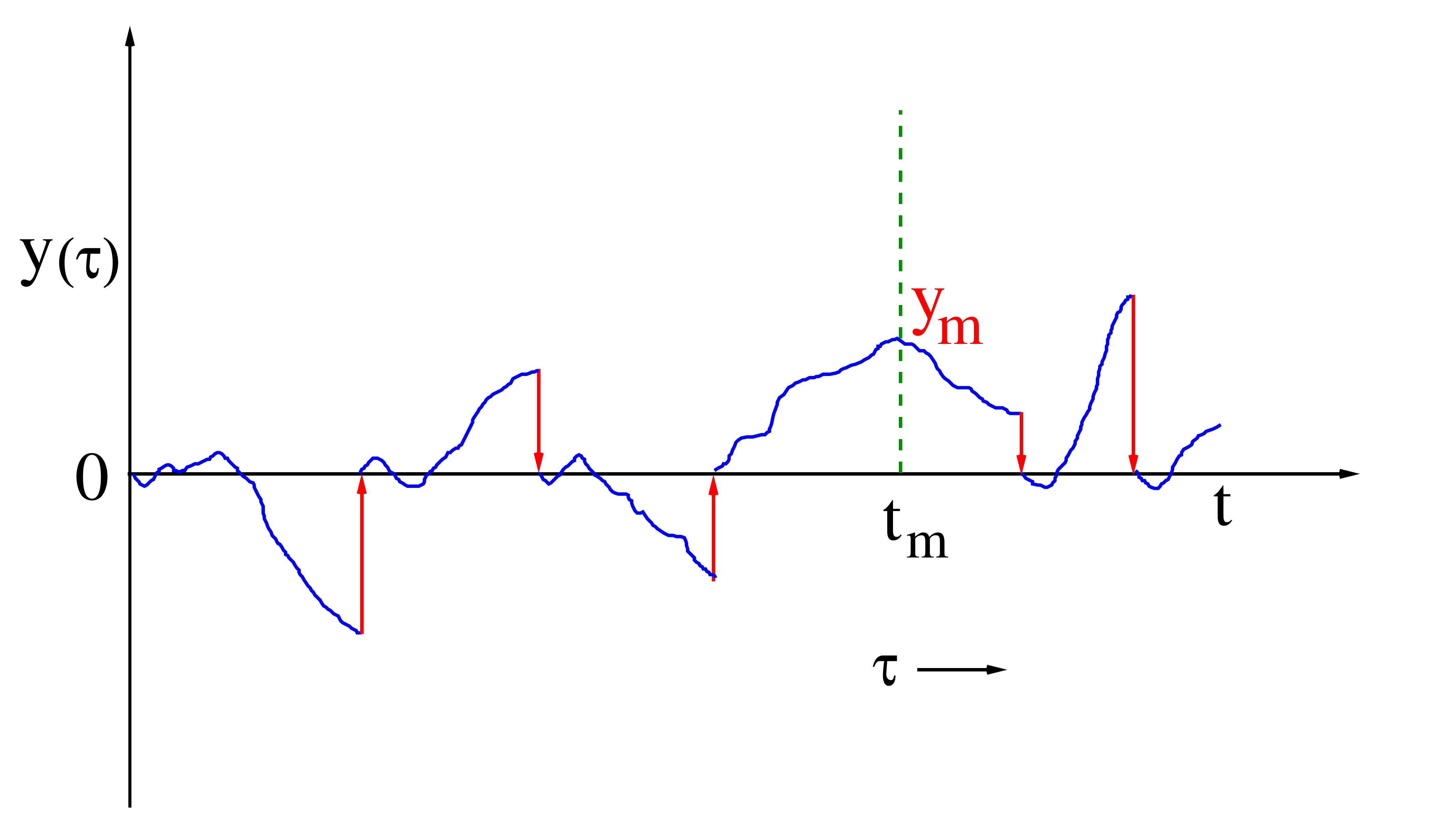}
        }
\caption{\label{fig_xy} A typical trajectory for the (a) $x$-component: $x(\tau)$ vs. $\tau$ for $\tau\in [0,t]$
and the (b) $y$ component: $y(\tau)$ vs. $\tau$ over the same interval $\tau\in [0,t]$, both starting at
the origin at $\tau=0$.
When a reset event happens (shown by (red) arrows), both $x$ and $y$ components are simultaneously reset to $0$.
Between resets, they evolve independently. The (red) dashed line indicates the maximum value $M$ of $x(\tau)$ in
the interval $[0,t]$ and it arrives at this maximum (actually at $x=M-\epsilon$ where $\epsilon$ is
a small cut-off (see text)) at time $t_m$. The value of the $y$ component at $\tau=t_m$ is denoted by
$y(\tau=t_m)= y_m$.}

\end{figure*}

We now switch on the resetting with a constant rate $r$ and our goal is to compute the second moment $\langle y_m^2\rangle$ of
the $y$ component at the time $t_m$ at which the $x$ component
achieves its maximum within the time interval $[0,t]$.
To compute this quantity, we first need to compute the probability
of a path of the process satisfying all the constraints. This
is done by using a path decomposition procedure that exploits the
Markov property of the process, as explained below.
This computation is best explained with the help of Fig.~\ref{fig_xy}, where we plot
schematically the trajectories of both the $x$ (left) and the $y$ component (right) in the time interval $[0,t]$, both
starting at $0$ at $\tau=0$. Let $t_m$ denote the time at
which $x(\tau)$ achieves its maximum value $M$ in $[0,t]$. This instant $t_m$ divides the full interval $[0,t]$ into two halves:
the left half during $[0,t_m]$ and the right half during $[t_m,t]$. On the left half, the $x$ component starts at the origin,
and has to stay below the level $M$ till $t_m$ and arrive at $M$ exactly at $t_m$. On the right half, the
process stays below the level $M$ during $[t_m, t]$. This restriction of staying below a given
level is usually implemented by putting an absorbing boundary condition at $x=M$. However, putting
this absorbing boundary at $x=M$ will
forbid the Brownian process to reach $x=M$ at $t_m$.
To circumvent this difficulty, we can put a small cut-off $\epsilon$ and say that the particle reaches the value $M-\epsilon$
at time $t_m$---then there is no problem in employing the absorbing boundary during the full interval $[0,t]$.
At the end of the calculation we will take the limit $\epsilon\to 0$ in an appropriate way.
While this procedure is perhaps not mathematically fully rigorous, it provides the quickest (and also a physically
transparent) way to
arrive at the correct final result. This limiting procedure using $\epsilon$ cut-off has been successfully used in
several examples of constrained Brownian motions before~\cite{MC_2005,MRKY_2008,SL_2010,PCMS_2013,PCMS_2015,MMS_2019,MMS_2020}.
Below we will use the same procedure for this problem also.

Let us first focus on the first interval $\tau\in [0,t_m]$. In this
interval, the $x$ component of the reset process, starting at the origin
and staying below $M$, reaches the position $M-\epsilon$ exactly at
$t_m$. The $y$ component, starting at the origin and undergoing
resettings at the same times as the $x$ process, reaches $y_m$ at
$\tau=t_m$. Note that the $y$ process is not restricted to be below $M$.
We then want to compute the propagator, i.e., the joint probability
density $G_r(x=M-\epsilon, y= y_m, \tau=t_m|M)$ for the two components to
reach respectively $M-\epsilon$ and $y_m$, with the former only staying
below the level $M$. The subscript $r$ in $G_r$ denotes the presence of
resetting to the origin with rate $r$. This propagator can again be
computed using the renewal method. We can write, as in Eq.
(\ref{renewal.1}) in the previous section,
\begin{equation}
G_r(M-\epsilon, y_m, t_m|M)= G_0(M-\epsilon, y_m, t_m|M)\, e^{- r\, t_m}
+ r\, \int_0^{t_m} dt_1\, e^{-r\, t_1}\, S_0(0,t_1|M)\, G_r(M-\epsilon,y_m,
t_m-t_1|M)\, ,
\label{Grt_renewal.1}
\end{equation}
where $S_0(0,t_1|M)$ denotes the survival probability of the $x$ component, i.e., a one dimensional
Brownian reset process up to time $t_1$, starting at the origin and with an absorbing boundary at $x=M$.
This is exactly the quantity whose Laplace transform we already computed in Eq.~(\ref{surv_x0.1}) in the previous section.
The quantity $G_0(M-\epsilon, y_m, t_m|M)$ is already computed
in Eq.~(\ref{G0_M.1}).
In Eq.~(\ref{Grt_renewal.1}), the first term represents no resetting
while the second term counts the contributions from at least one resetting. For this we consider
the first resetting at $t_1$ that occurs before
$t_m$, and after that the process renews. During $[0,t_1]$ the $x$ process stays below $M$ which is
ensured by the factor $S_0(0,t_1|M)$ inside the integral on the rhs of Eq.~(\ref{Grt_renewal.1}).
Note that for the $y$ process we have no restriction whatsoever.
The factor
$G_r(M-\epsilon,y_m,t_m-t_1|M)$ ensures that the renewed $(x,y)$
process starting at $t_1$ from the origin
arrives at $(M-\epsilon, y_m)$ after a time $t_m-t_1$, with the
$x$ component staying below $M$ during $t_m-t_1$. We take the product and integrate over $t_1$ from
$0$ to $t_m$. Again, the convolution structure of Eq.~(\ref{Grt_renewal.1}) signals that it simplifies in the
Laplace space (with respect to time). Indeed, defining
${\tilde G}_r(x,y,s|M)= \int_0^{\infty} G_r(x,y,t_m|M)\, e^{-s\, t_m}\, dt_m$
and taking the Laplace transform of Eq.~(\ref{Grt_renewal.1}) with respect to $t_m$ gives
\begin{equation}
{\tilde G}_r(M-\epsilon, y_m, s|M)=
\frac{{\tilde G}_0(M-\epsilon, y_m, r+s|M)}{1- r\, {\tilde S}_0(0,r+s)|M)}
= \frac{(r+s)\, {\tilde G}_0(M-\epsilon, y_m, r+s|M)}{s+r\,
e^{-\sqrt{\frac{r+s}{D}}\, M}}\, ,
\label{Grs_renewal.1}
\end{equation}
where we used the expression for ${\tilde S}_0(0,r+s)|M)$ from
Eq.~(\ref{surv_x0.1}). The relation in Eq.~(\ref{Grs_renewal.1})
expresses the constrained Green's function of the process
with resetting in terms of
the constrained Green's function without resetting $G_0$
computed in Eq.~(\ref{G0_M.1}).

We now focus on the second time interval $\tau\in [t_m,t]$. This is easy
because during this interval we just have to ensure that only the $x$
process, starting at $M-\epsilon$ at time $t_m$, stays below the
level $M$ during the interval $[t_m,t]$. For the $y$ process we have
no restriction. Hence this probability is simply
$S_r(M-\epsilon,t-t_m|M)$, i.e., the survival probability for the $x$
process during the interval $[t_m,t]$, starting at $M-\epsilon$ at time $t_m$, with an absorbing boundary at $M$.
The Laplace transform of this probability
has already been computed in Eq.~(\ref{srx0.2}).

The total probability of such a constrained path of the joint $x$ and $y$
processes is then proportional to the product of the probabilities
in the first ($[0,t_m]$) and the second ($[t_m,t]$) time intervals--this
follows from the Markov property of the process. We denote this total
probability by $P_r(M,t_m,y_m|t,\epsilon)$ where $M$, $t_m$ and $y_m$
are random variables, while $t$ and $\epsilon$ are fixed parameters.
Taking this product gives
\begin{equation}
P_r(M,t_m,y_m|t,\epsilon)= {\cal N}(\epsilon)\, G_r(M-\epsilon,y_m,t_m|M)\,
S_r(M-\epsilon,t-t_m|M)\, ,
\label{Pr_joint.1}
\end{equation}
where ${\cal N}(\epsilon)$ is just a normalization constant, such that
the joint probability density $P_r(M,t_m,y_m|t,\epsilon)$ is
normalized to unity when integrated over $M$, $t_m$ and $y_m$. In principle, ${\cal N}(\epsilon)$
can also depend on $t$. However here, we assume and verify a posteriori that it is indeed independent
of $t$.

Let us first integrate over $t_m$. This gives the joint probability
density of $M$ and $y_m$
\begin{equation}
P_r(M,y_m|t,\epsilon)= {\cal N}(\epsilon)\, \int_0^{t} G_r(M-\epsilon,y_m,t_m|M)\,
S_r(M-\epsilon,t-t_m|M)\, dt_m\, .
\label{Pr_Mym.1}
\end{equation}
It is convenient to take the Laplace transform of Eq.~(\ref{Pr_Mym.1})
with respect to $t$ to get
\begin{equation}
{\tilde P}_r(M,y_m|s,\epsilon)= \int_0^{\infty} P_r(M,y_m|t,\epsilon)\,
e^{-s\,t}\, dt= {\cal N}(\epsilon)\, {\tilde G}_r(M-\epsilon,y_m,s|M) \,
{\tilde S}_r(M-\epsilon,s|M)\, .
\label{Pr_Mym_lap.1}
\end{equation}
We now use the explicit expressions for ${\tilde G}_r(M-\epsilon,y_m,s|M)$
from Eq.~(\ref{Grs_renewal.1}) and ${\tilde S}_r(M-\epsilon,s|M)$
from Eq.~(\ref{srx0.2}) to get
\begin{equation}
{\tilde P}_r(M,y_m|s,\epsilon)={\cal N}(\epsilon)\, \frac{(r+s)\,
\left(1-e^{-\sqrt{\frac{r+s}{D}}\, \epsilon}\right)}{
\left(s+r\, e^{-\sqrt{\frac{r+s}{D}}\, M}\right)^2}\,
{\tilde G}_0(M-\epsilon,y_m,r+s|M)\, .
\label{Pr_Mym_lap.2}
\end{equation}
Using the expression for $G_0$ from Eq.~(\ref{G0_M.1}) we then obtain
\begin{equation}
{\tilde P}_r(M,y_m|s,\epsilon)={\cal N}(\epsilon)\, \frac{(r+s)\,
\left(1-e^{-\sqrt{\frac{r+s}{D}}\, \epsilon}\right)}{
\left(s+r\, e^{-\sqrt{\frac{r+s}{D}}\, M}\right)^2}\,
\int_0^{\infty} \frac{dT}{4\,\pi\,D\, T}\, e^{-(r+s)\,T}\,
e^{-\frac{y_m^2}{4\,D\,T}}\, \left[ e^{-\frac{(M-\epsilon)^2}{4\,D\,T}}
- e^{-\frac{(M+\epsilon)^2}{4\,D\,T}}\right]\, .
\label{Pr_Mym_lap.3}
\end{equation}
We now take the $\epsilon\to 0$ limit. To leading order in $\epsilon$ it gives
\begin{equation}
{\tilde P}_r(M,y_m|s,\epsilon\to 0)= \left[\lim_{\epsilon\to 0}
{\cal N}(\epsilon)\, \epsilon^2\right]\,
\frac{(r+s)^{3/2}}{D^{3/2}\, \left(s+r\, e^{-\sqrt{\frac{r+s}{D}}\, M}\right)^2}\,
\int_0^{\infty} \frac{dT}{4\,\pi\,D\, T}\, e^{-(r+s)\,T}\,
e^{-\frac{y_m^2}{4\,D\,T}}\, \frac{M}{T}\, e^{-\frac{M^2}{4\, D\, T}}\, .
\label{Pr_Mym_lap.4}
\end{equation}

To determine the normalization constant, we integrate
Eq.~(\ref{Pr_Mym_lap.4}) over $y_m$ to get the Laplace transform
of the marginal distribution of $M$
\begin{eqnarray}
{\tilde P}_r(M,s)=
\int_{-\infty}^{\infty} {\tilde P}_r(M,y_m|s,\epsilon\to 0)\, dy_m
&=& \left[\lim_{\epsilon\to 0}{\cal N}(\epsilon)\, \epsilon^2\right]\,
\frac{(r+s)^{3/2}}{D^{3/2}\, \left(s+r\, e^{-\sqrt{\frac{r+s}{D}}\, M}\right)^2}\,
\frac{M}{\sqrt{4\,\pi\,D}}\,\int_0^{\infty} \frac{dT}{T^{3/2}}\,
e^{-(r+s)\,T- \frac{M^2}{4\,D\,T}}\, \nonumber \\
&=& \left[\lim_{\epsilon\to 0}{\cal N}(\epsilon)\, \epsilon^2\right]\,
\frac{(r+s)^{3/2}}{D^{3/2}\, \left(s+r\, e^{-\sqrt{\frac{r+s}{D}}\, M}\right)^2}\,
e^{-\sqrt{\frac{r+s}{D}}\, M}\, ,
\label{Pr_M_lap.1}
\end{eqnarray}
where the integral over $T$ is performed explicitly using the
identity $\int_0^{\infty} dT\, T^{-3/2}\, e^{-a T- b/T}=
\sqrt{\frac{\pi}{b}}\, e^{-2\,\sqrt{a\,b}}$ for $a$ and $b$ positive.
However, ${\tilde P}_r(M,s)$ was already computed in Eq.~(\ref{pdf_maxr.1}).
Hence, comparing the rhs of Eqs.~(\ref{Pr_M_lap.1}) and (\ref{pdf_maxr.1})
we immediately get the normalization constant
\begin{equation}
\lim_{\epsilon\to 0} {\cal N}(\epsilon)\, \epsilon^2= D \, .
\label{norm.1}
\end{equation}
Using this normalization constant in Eq.~(\ref{Pr_Mym_lap.4}) finally
gives the Laplace transform of the joint distribution of $M$ and $y_m$
\begin{equation}
{\tilde P}_r(M,y_m,s)= {\tilde P}_r(M,y_m|s,\epsilon\to 0)=
\frac{M\,
(r+s)^{3/2}}{{D}^{3/2}\, \left(s+r\, e^{-\sqrt{\frac{r+s}{D}}\, M}\right)^2}\,
\int_0^{\infty} \frac{dT}{4\,\pi\, T^2}\,
e^{-(r+s)\,T- \frac{(M^2+y_m^2)}{4\,D\,T}}\, .
\label{Mym_joint.1}
\end{equation}
From this joint distribution, we can then compute the Laplace transform
of $\langle y_m^2\rangle(t)$ as
\begin{equation}
\int_0^{\infty} \langle y_m^2\rangle(t)\, e^{-s\, t}\, dt
= \int_0^{\infty} dM\, \int_{-\infty}^{\infty} dy_m\, y_m^2\, {\tilde P}_r(M,y_m,s)\, .
\label{ym2_lap.1}
\end{equation}
Substituting Eq.~(\ref{Mym_joint.1}) on the rhs of Eq.~(\ref{ym2_lap.1})
and carrying out the integrals over $y_m$ and $T$ explicitly we finally
get a relatively simple expression
\begin{equation}
\int_0^{\infty} \langle y_m^2\rangle(t)\, e^{-s\, t}\, dt
= D\, \int_0^{\infty} dm\, \frac{ m\, e^{-m}}{\left(s+r\, e^{-m}\right)^2}\, ,
\label{ym2_lap.2}
\end{equation}
where we made a change of variable $m= \sqrt{(r+s)/D}\, M$ in the integral
over $M$. We can now invert the Laplace transform in Eq.~(\ref{ym2_lap.2}) using
${\cal L}_s^{-1}\left[(s+a)^{-2}\right]= t\, e^{-a\, t}$. This gives
our final result in this section
\begin{equation}
\langle y_m^2\rangle(t)= D\, t\, \int_0^{\infty} dm\, m\, e^{-m- r\,t\, e^{-m}}
= \frac{2\,D}{r}\, V(r\, t)\,
\label{ym2t.1}
\end{equation}
with the scaling function $V(z)$ given by
\begin{equation}
V(z)= \frac{z}{2}\,\int_0^{\infty} dm\, m\, e^{-m- z\, e^{-m}}=
\frac{1}{2}\,(\Gamma[0,z]+\ln z+\gamma_E)\, ,
\label{Vz.1}
\end{equation}
where $\Gamma[0,z]=\int_z^{\infty} \frac{e^{-t}}{t}\, dt$.
It has the asymptotics behaviors
\begin{eqnarray}
V(z) \approx \begin{cases}
\frac{z}{2}- \frac{z^2}{8} + O(z^3)  \quad\quad\quad\quad\, {\rm as}\quad z\to 0 \;, \\
\\
\frac{1}{2}\,(\ln z + \gamma_E) +O(1/z)  \quad\, {\rm as}\quad z\to \infty \;.
\end{cases}
\label{Vz_asymp}
\end{eqnarray}

\section{The mean perimeter and the mean area of the convex hull}\label{Sec:final}

With all the basic ingredients derived in the previous two sections,
we then go on to compute the
mean perimeter and the mean area of the convex hull of the $2$-d Brownian
motion with resetting to the origin. As discussed in Section-II, the
mean perimeter of the convex hull of an isotropic $2$-d stochastic process
is given by $\langle L(t)\rangle = 2\,\pi\, \langle M(t)\rangle$, where
$\langle M(t)\rangle$ is the expected maximum of the one dimensional $x$
component process up to time $t$. Using the result from Eq.~(\ref{ex_max.4})
we then obtain the exact result for the mean perimeter of the convex hull
of the $2$-d RBM for all $t$
\begin{equation}
\langle L(t)\rangle = 2\,\pi\, \sqrt{\frac{D}{r}}\, f_1(r\,t)
\label{meanp_1}
\end{equation}
where the scaling function $f_1(z)$ is given in Eq.~(\ref{ex_max.4})
with asymptotic behaviors in Eq.~(\ref{f1z_asymp}). Using the asymptotic behavior of $f_1(z)$
in the limit $z \to 0$, we find that for $t \ll 1/r$, $\langle L(t) \approx \sqrt{16 \pi D t}\rangle$, thus recovering
the Brownian result in Eq. (\ref{Lt_BM}), as expected. In contrast, for $t \gg 1/r$, the mean perimeter
grows logarithmically  
\begin{equation}
\langle L(t)\rangle \approx  2\, \pi\, \sqrt{\frac{D}{r}}\, \ln(rt) \, . 
\label{Lt_asymp.1}
\end{equation}

\begin{figure}[htb]
    \centering
    \subfigure[\label{fig_perimeter}]{
        \includegraphics[scale=1]{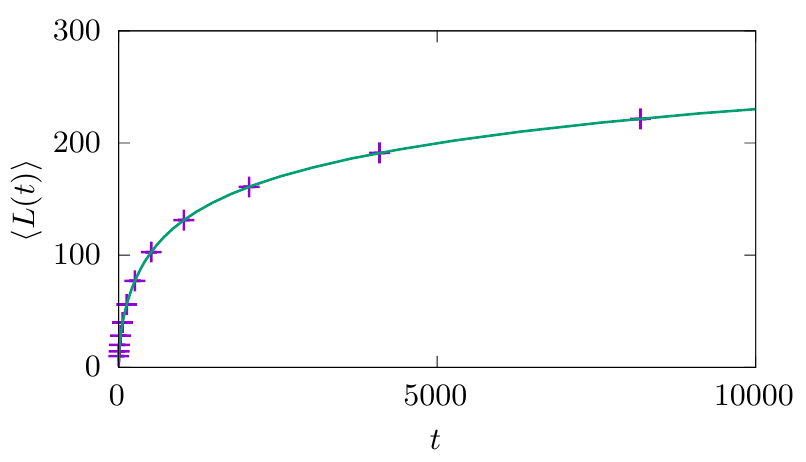}
    }
    \subfigure[\label{fig_area}]{
        \includegraphics[scale=1]{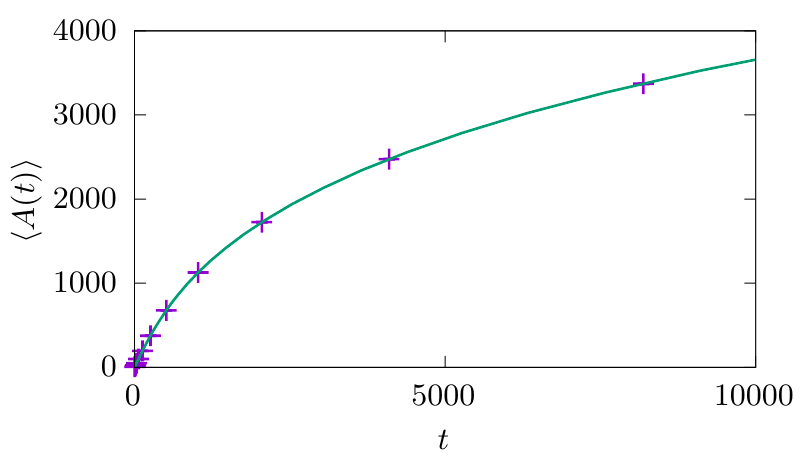}
    }
    \caption{\label{fig_numeric}
        The mean \subref{fig_perimeter} perimeter $\langle L(t)\rangle$
        and \subref{fig_area} area $\langle A(t)\rangle$ plotted as a function of time $t$.
        The solid lines represent the exact result
        \subref{fig_perimeter} $\langle L(t)\rangle = 2\,\pi\, \sqrt{\frac{D}{r}}\, f_1(r\,t)$
        given in Eq.~(\ref{meanp_1})
        and \subref{fig_area} $\langle A(t)\rangle=
        (2D/r)\, f_2(rt)$ with $f_2(z)$ given in Eq.~(\ref{Qz.1}).
        The symbols represent the respective results from numerical simulations with
        errorbars smaller than the lines. {Here $D=1/2$ and $r=10^{-2}$.}
    }
\end{figure}

Our numerical simulations show perfect agreement with our
analytical prediction for all $t$ (see Fig.~\ref{fig_numeric}\subref{fig_perimeter}).

The mean area of the convex hull is given by (see Section-II)
\begin{equation}
\langle A(t)\rangle= \pi\, \left[\langle M^2(t)\rangle-
\langle y_m^2\rangle (t)\right]\, .
\label{meana.1}
\end{equation}
Using the explicit results for $\langle M^2(t)\rangle$ from Eq.~(\ref{Hz_def})
and for $\langle y_m^2\rangle (t)$ from Eq.~(\ref{ym2t.1}) we get,
for all $t$,
\begin{equation}
\langle A(t)\rangle=\frac{2\pi D}{r}\, f_2(rt)\, ,
\label{area.2}
\end{equation}
where the scaling function $f_2(z)$ is given by
\begin{equation}
f_2(z)= H(z)-V(z)= z\,
\, _3F_3\left[\{1,1,1\},\{2,2,2\},-z\right]- \frac{1}{2} \left[\Gamma[0,z]+
\ln z + \gamma_E\right]\, .
\label{Qz.1}
\end{equation}
It has the following asymptotic behaviors
\begin{eqnarray}
f_2(z) \approx \begin{cases}
& \frac{z}{2}- \frac{z^3}{108}+ +\frac{z^4}{384}
\quad\, {\rm as}\quad z\to 0 \\
\\
& \frac{1}{2}\left[\ln^2 z+ \left(2\gamma_E-1\right)\, \ln z +
\gamma_E^2-\gamma_E+\frac{\pi^2}{6}\right]+
O(1/z) \quad {\rm as}\quad z\to \infty
\end{cases}
\label{Qz_asymp}
\end{eqnarray}
Note that when $t\ll 1/r$, using $f_2(z) \approx z/2$ as $z\to 0$, we recover
the standard Brownian motion result $\langle A(t)\rangle \approx \pi\, D\, t$ in Eq. (\ref{At_BM}).
In contrast, for large $t\gg 1/r$, the mean area of the convex hull
grows as 
\begin{equation}
\langle A(t) \rangle \approx \pi\, \frac{D}{r}\, \ln^2(rt) \, .
\label{At_asymp.1}
\end{equation}
In Fig.~\ref{fig_numeric}\subref{fig_area}, we compare
our analytical prediction with numerical simulations, finding
excellent agreement for all $t$.

\begin{figure}[t]
\includegraphics[width=0.7\linewidth]{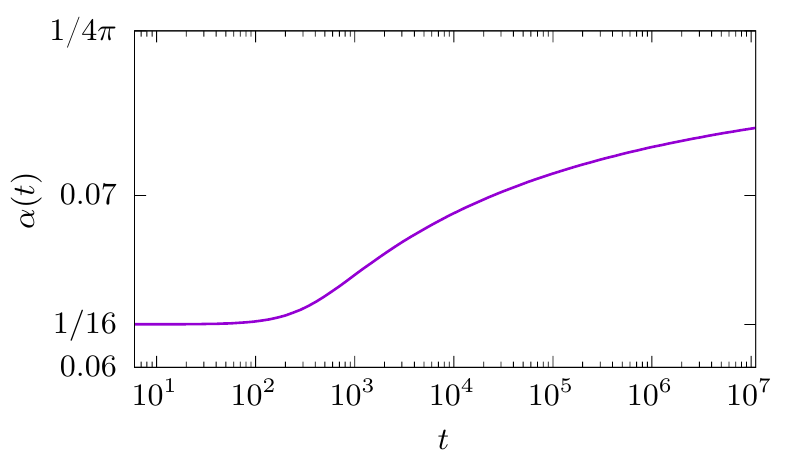}
\caption{{Plot of $\alpha(t) ={\langle A(t)\rangle}/{[\langle L(t) \rangle]^2}$, given in Eq. (\ref{alphat.1}) together with Eqs. (\ref{ex_max.4}) and (\ref{Qz.1}), as a function of $t$, on a log-log scale, for $D=1/2$ and $r= 10^{-2}$. When $t\to \infty$, $\alpha(t) \to 1/(4\pi)$, which shows that the convex hull reaches a circular shape. Note however that this convergence to a circular shape is quite slow [from Eq. (\ref{Fz_asymp}), one sees that $1/(4 \pi) - \alpha(t) \propto 1/(\ln (rt))$, as $t \to \infty$].}} \label{Fig_plot_alpha}
\end{figure}

{Let us point out an interesting geometric fact that emerged from our exact results.
For a perfect circle of radius $R$, the perimeter is $L=2\pi R$ while the area is $A=\pi R^2$.
Hence the dimensionless ratio
\begin{equation}
\alpha=\frac{A}{L^2}= \frac{1}{4\pi}\, .
\label{alpha_def}
\end{equation}
The isoperimetric inequality says that for any arbitrary shape in two dimensions 
$\alpha\le 1/(4\pi)$,
i.e, the isoperimetric upper bound is saturated by the circular shape. 
The closer the value of $\alpha$
to this upper bound $1/(4\pi)$, the more circular is the shape of the domain.  
Now, for the convex hull of a planar RBM, our exact results
in Eqs. (\ref{meanp_1}) and (\ref{area.2}) show that the ratio  
\begin{equation}
\alpha(t)= \frac{\langle A(t)\rangle}{[\langle L(t) \rangle]^2}= F(rt)\, ;
\quad\,\, {\rm where}\quad\, F(z)= \frac{1}{2\pi}\,\frac{f_2(z)}{f_1^2(z)}\, 
\label{alphat.1}
\end{equation}
valid at any time $t$, with the scaling functions $f_1(z)$ and $f_2(z)$ given
respectively in Eqs. (\ref{ex_max.4}) and (\ref{Qz.1}). 
The function $F(z)$ has the 
asymptotic behaviors
\begin{eqnarray}
F(z) \approx \begin{cases}
& \frac{1}{16} \left(1+ \frac{1}{270}z^2 \right)\quad\, {\rm as}\quad z\to 0 \\
\\
& \frac{1}{4\pi} \left(1 - \frac{1}{\ln z}\right)\quad\quad {\rm as} \quad z\to \infty \;.
\end{cases}
\label{Fz_asymp}
\end{eqnarray}
Hence, for $t\ll 1/r$, when the RBM effectively behaves like a Brownian motion without resetting,
the ratio $\alpha(t)\approx 1/16 < 1/(4\pi)$, indicating that at short times the shape of the
convex hull is far from circular. In contrast, for $t\gg 1/r$, $\alpha(t)\approx 1/(4\pi)$
indicating that repeated resettings drive the convex hull of the RBM to a circular shape
at late times. The exact function $\alpha(t)$ in Eq. (\ref{alphat.1}) precisely describes
the evolution of the shape of the convex hull from a non-circular to circular shape as
$t\to \infty$.} {In Fig. \ref{Fig_plot_alpha} we show a plot of $\alpha(t)$,  given in Eq. (\ref{alphat.1}) together with Eqs. (\ref{ex_max.4}) and (\ref{Qz.1}), as a function of $t$.}

To simulate RBM of duration $t$, we discretized the duration into
intervals of length $\Delta\tau$ and follow the rules of Eq.~\eqref{RBM_rules}.
That is, we reset for each interval the position with probability $r\Delta\tau$
with $r=0.01$ to the origin and add a
a Gaussian distributed jump with standard deviation $\sigma = \sqrt{\Delta\tau}$,
corresponding to $D=1/2$, to the current position. The approximation naturally
becomes better with smaller values for $\Delta\tau$; here we chose $\Delta\tau = t/10^6$.
The positions visited at the end of each interval are used to calculate the
convex hull using Andrew's monotone chain algorithm \cite{Andrew1979Another} after
pruning the point set with Akl's heurisitic \cite{Akl1978Fast}. To calculate
the averages of their perimeter and area we generated $10^5$ independent
realizations per value of $t$, leading to relative statistical errors in the order of a few
permil, within which they are always compatible with our analytical predictions.

\section{Conclusion}\label{Sec:conclusion}

In this paper, we have obtained exact formulae for the mean perimeter and the mean
area of the convex hull of a $2$-d RBM of fixed duration $t$. Our formulae are valid for all
time $t$. For time $t \ll 1/r$, we recover the well known Brownian results in the absence of
resetting. For $r>0$, our results show that, at late times $t \gg 1/r$, the mean perimeter and
the mean area grow respectively as $\ln(rt)$ and $\ln^2(rt)$. Our main conclusion is
thus that, even though the position distribution becomes stationary at late times, the convex hull
keeps growing, albeit logarithmically slowly. {Moreover, our exact results for the
mean perimeter and the mean area indicate 
that in the presence of resetting, the
convex hull of a $2$-d RBM approaches a circular shape at late times, as indicated by
the saturation of the isoperimetric upper bound.}
It would be interesting to compute the full distribution of the
perimeter and the area, though we do not have any analytical method currently to go beyond
the first moments. In fact, even for standard Brownian motion without resetting, these distributions
are only known numerically \cite{Claussen2015Convex,schawe2017highdim,SHM2018,schawe2019true}.

In this computation, we have assumed that the resetting occurs instantaneously. However, in reality,
when the animal comes back to its nest, it takes some time to come back -- this is usually referred
to as the refractory period when it is not actively searching for food. The effects of such refractory period
on the resetting process have been studied in one-dimensional models \cite{Reuveni_16,HK2016,EM2019,BS2020}. It would be interesting to
see how the statistics of the convex hull depends on the refractory periods.

\appendix

\section{Laplace inversion of Eq.~(\ref{lt_max.1})}

The Laplace transform in Eq.~(\ref{lt_max.1}) can be inverted using the convolution theorem as follows.
We denote
\begin{eqnarray}
{\cal L}_s^{-1}\left [ \ln \left(\frac{r+s}{s}\right)\right]&= & g_1(t) \label{g1_def}\\
{\cal L}_s^{-1}\left [ \frac{\sqrt{r+s}}{s}\right] &=& g_2(t)  \label{g2_def}
\end{eqnarray}
and then apply the convolution theorem to
Eq.~(\ref{lt_max.1}) to get
\begin{equation}
\langle M(t)\rangle= \frac{ \sqrt{D}}{r}\, \int_0^t g_1(\tau)\, g_2(t-\tau)\, d\tau \, .
\label{convol.1}
\end{equation}
Hence we need just to find the inverse functions $g_1(t)$ and $g_2(t)$.

To find $g_1(t)$, we note the following simple identity
\begin{equation}
\int_0^{\infty} \frac{1-e^{-rt}}{t}\,e^{-st} dt= \ln\left(\frac{r+s}{s}\right)\, ,
\label{g1t.1}
\end{equation}
indicating that
\begin{equation}
g_1(t)= \frac{1- e^{-rt}}{t}\, .
\label{g1t.2}
\end{equation}

Next we need to find $g_2(t)$ from Eq.~(\ref{g2_def}). We rewrite it as
\begin{equation}
g_2(t)= {\cal L}_s^{-1}\left [ \frac{\sqrt{r+s}}{s}\right]=
{\cal L}_s^{-1}\left [\frac{r+s}{s\sqrt{r+s}}\right]= {\cal L}_s^{-1}\left [ \frac{1}{\sqrt{r+s}}+
\frac{r}{s\,\sqrt{r+s}} \right]\, .
\label{g2t.1}
\end{equation}
The first term on the rhs can be immediately inverted
\begin{equation}
{\cal L}_s^{-1}\left [ \frac{1}{\sqrt{r+s}}\right]= \frac{e^{-r t}}{\sqrt{\pi\, t}}\, .
\label{g2t.2}
\end{equation}
The second term can be inverted by noting ${\cal L}_s^{-1}[1/s]=1$
and then using the convolution theorem where we make use of Eq.~(\ref{g2t.2}). This gives
\begin{equation}
{\cal L}_s^{-1}\left [ \frac{r}{s\sqrt{r+s}} \right]= r\, \int_0^t \frac{e^{-r \tau}}{\sqrt{\pi\, \tau}}\, d\tau=
\sqrt{r}\, {\rm erf}\left(\sqrt{r\, t}\right)\, .
\label{g2t.3}
\end{equation}
Adding Eqs.~(\ref{g2t.2}) and (\ref{g2t.3}) gives
\begin{equation}
g_2(t)= \sqrt{r}\, \left[\frac{e^{-r t}}{\sqrt{\pi\, r\, t}} + {\rm erf}\left(\sqrt{r\, t}\right)\right]\, .
\label{g2t_final}
\end{equation}
Substituting $g_1(t)$ and $g_2(t)$ in Eq.~(\ref{convol.1}) and changing the variable $r\,\tau=y$ in the convolution
integral gives the result in
Eq.~(\ref{ex_max.4}). We note that while this inversion was already mentioned
in Ref.~\cite{EM2011.1}, the details of the inversion was not provided there.
We include it here for the sake of completeness.

\end{document}